\documentclass[aps,prl,twocolumn,superscriptaddress,floats]{revtex4-1}
\usepackage{txfonts}
\usepackage{amssymb}
\usepackage{graphicx}
\usepackage{dcolumn}
\usepackage{bm}
\usepackage[colorlinks,linkcolor=blue,anchorcolor=blue,citecolor=blue,urlcolor=blue]{hyperref}
\usepackage{hyperref}

\begin{document}

\title{ Neutron-Diffraction Measurements of an Antiferromagnetic Semiconducting Phase\\in the Vicinity of the High-Temperature Superconducting State of K$_x$Fe$_{2-y}$Se$_2$
}
\author{Jun Zhao}
\email{zhaoj@fudan.edu.cn}
\affiliation{
Department of Physics, University of California, Berkeley, California 94720, USA
}
\affiliation{
Miller Institute for Basic Research in Science, Berkeley, California 94720, USA
}

\author{Huibo Cao}
\affiliation{
Quantum Condensed Matter Division, Oak Ridge National Laboratory, Oak Ridge, TN 37831
}
\author{E. Bourret-Courchesne}
\affiliation{
Materials Sciences Division, Lawrence Berkeley National Laboratory, Berkeley, California 94720, USA
}

\author{D. -H. Lee}
\affiliation{
Department of Physics, University of California, Berkeley, California 94720, USA
}

\affiliation{
Materials Sciences Division, Lawrence Berkeley National Laboratory, Berkeley, California 94720, USA
}

\author{R. J. Birgeneau}
\affiliation{
Department of Physics, University of California, Berkeley, California 94720, USA
}
\affiliation{
Department of Materials Science and Engineering, University of California, Berkeley, California 94720, USA
}

\begin{abstract}

The recently discovered K-Fe-Se high temperature superconductor has caused heated debate regarding the nature of its parent compound. Transport, angle-resolved photoemission spectroscopy, and STM measurements have suggested that its parent compound could be insulating, semiconducting or even metallic [M. H. Fang, H.-D. Wang, C.-H. Dong, Z.-J. Li, C.-M. Feng, J. Chen, and H. Q. Yuan, \href{http://dx.doi.org/10.1209/0295-5075/94/27009}{Europhys. Lett. \textbf{94}, 27009 (2011)}; F. Chen \textit{et al.}, \href{http://dx.doi.org/10.1103/PhysRevX.1.021020}{Phys. Rev. X \textbf{1}, 021020 (2011)}; and W. Li \textit{et al.}, \href{http://dx.doi.org/10.1103/PhysRevLett.109.057003}{Phys. Rev. Lett. \textbf{109}, 057003 (2012)}]. Because the magnetic ground states associated with these different phases have not yet been identified and the relationship between magnetism and superconductivity is not fully understood, the real parent compound of this system remains elusive. Here, we report neutron-diffraction experiments that reveal a semiconducting antiferromagnetic (AFM) phase with rhombus iron vacancy order. The magnetic order of the semiconducting phase is the same as the stripe AFM order of the iron pnictide parent compounds. Moreover, while the $\sqrt{5} \times \sqrt{5}$ block AFM phase coexists with superconductivity, the stripe AFM order is suppressed by it. This leads us to conjecture that the new semiconducting magnetic ordered phase is the true parent phase of this superconductor.
\end{abstract}


\maketitle

Identifying the parent compound of a high-temperature superconductor and understanding its magnetic and structural properties are important because the fluctuating version of the ordered magnetic mode can be the trigger of Cooper pairing \cite{lee,johnston}. In the parent compounds of the cuprate superconductors, it is well established that the antiferromagnetic (AFM) order arises from superexchange interactions between local moments driven by strong electron correlations \cite{lee}. However, the origin of the magnetism in the iron pnictides is still a matter of debate. It is widely believed that the magnetism is related to Fermi surface scattering, since the AFM order is characterized by a wave vector connecting centers of the electron and hole Fermi surfaces \cite{johnston}. Alternatively, a local moment picture has also been proposed for the iron pnictides, similar to that of the cuprates \cite{hu,si2,kotliar,xu}. The recently discovered alkali-metal-intercalated iron selenide superconductors A$_x$Fe$_{2-y}$Se$_2$ (A=K, Rb, Cs) has attracted great interest because they display a variety of properties unprecedented for the cuprates and iron pnictides \cite{chenxl,mazin,fangmh}. For example, angle-resolved photoemission spectroscopy (ARPES) measurements have suggested that the normal state Fermi surface of the superconducting compound does not have any hole pockets \cite{donglai,zhou,ding}. In terms of their magnetic properties, the A$_x$Fe$_{2-y}$Se$_2$ superconductors are equally unusual. Early work suggested that a strong $\sqrt{5} \times \sqrt{5}$ block AFM order coexists with superconductivity \cite{weibao}, while, recently, growing evidence suggests that the block AFM phase is spatially separated from the superconducting phase \cite{donglai1,schermandini2,ksenofontov,nanlin,ricci,charnukha} and that superconductivity only exists in a minority phase with no block AFM order \cite{li2,wangyy,li,texier,borisenko}. In addition, ARPES and transport measurements have identified several different phases in the K$_x$Fe$_{2-y}$Se$_2$ compounds: an insulating phase with a 500 meV band gap (from now on, we refer to it as the ``insulating" phase), a small gap insulating phase with a 40 meV gap (from now on, we refer to it as the ``semiconducting" phase), and a superconducting phase with only electronlike Fermi surfaces \cite{donglai1}. The same measurements also suggested that the band structure of the superconducting phase was more similar to that of the semiconducting phase rather than that of the insulating phase \cite{donglai1}. This, plus our finding that the semiconducting AFM phase and superconductivity mutually suppress each other, leads us to conjecture that the new phase we shall report in the following is the parent state of superconductivity. However, more detailed studies of how the magnetic excitation evolves from the semiconducting AFM phase to the superconducting phase is needed to confirm this conjecture and gain deeper insight into the relationship between the magnetism in this phase and Cooper pairing.

\begin{figure}[t]
\centering\includegraphics[width=0.5\textwidth]{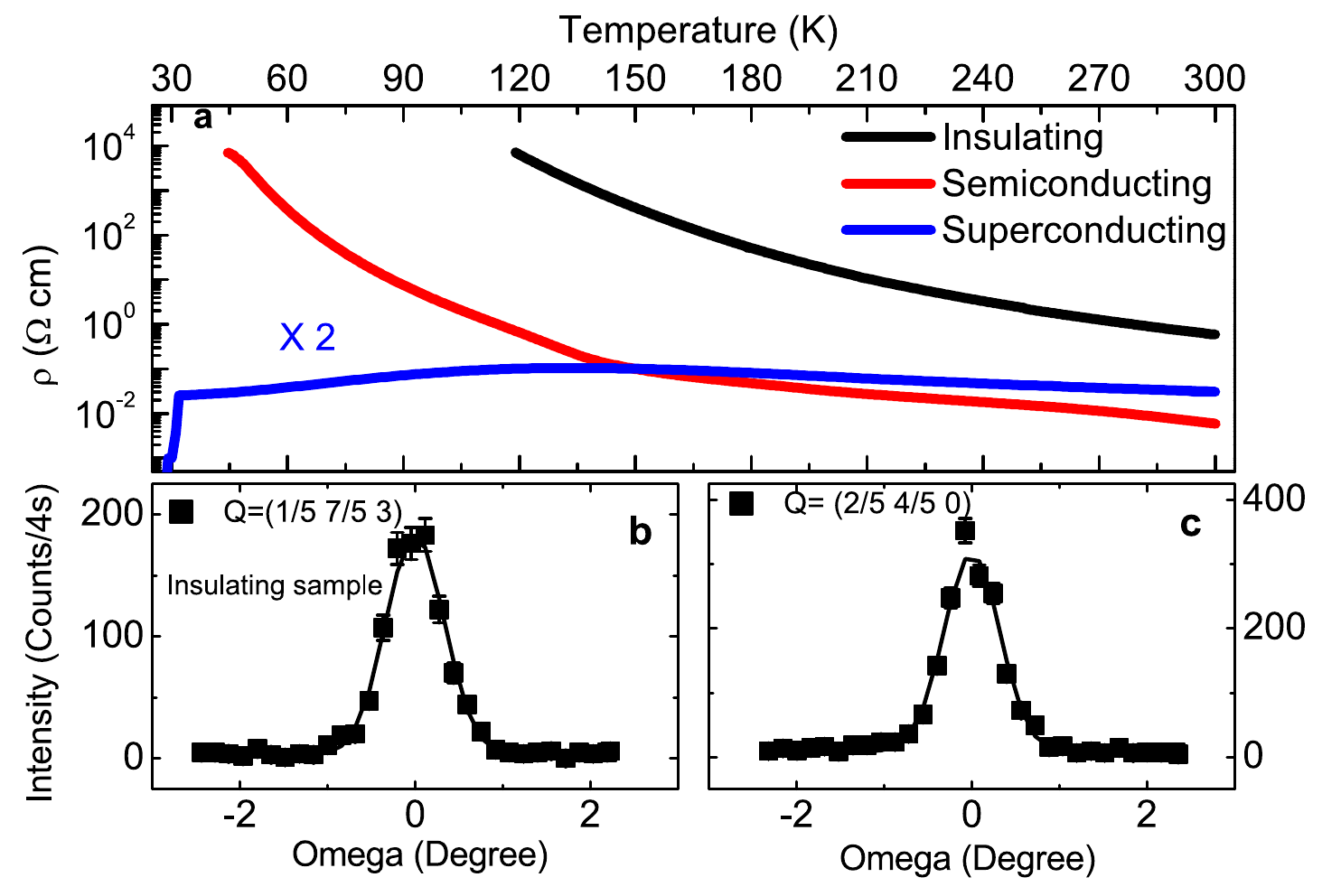}
\caption{\label{fig1}Resistivity of various K$_x$Fe$_{2-y}$Se$_2$ samples and $\sqrt{5} \times \sqrt{5}$ superstructure reflections of the insulating K$_{0.8}$Fe$_{1.6}$Se$_2$. The Bragg reflections of insulating K$_{0.8}$Fe$_{1.6}$Se$_2$ can be fully described with $\sqrt{5} \times \sqrt{5}$ block AFM order and iron vacancy order. (a) In-plane resistivity as a function of temperature for the insulating sample, the semiconducting sample, and the superconducting sample with $T_c$= $30$ K. The resistivity of the superconducting sample is multiplied by 2. (b) Magnetic reflection associated with $\sqrt{5} \times \sqrt{5}$ block AFM order of the insulating sample. (c) Nuclear reflection associated with $\sqrt{5} \times \sqrt{5}$ iron vacancy order of the insulating sample. The lattice parameters of the insulating sample are $a$=$b$=$5.504(3)$ $\AA$, $c$=$14.03(6)$ $\AA$ at 5 K in the orthorhombic notation. The error bars indicate 1 standard deviation throughout the letter.
}
\end{figure}

We use neutron diffraction to study the structure, magnetic order and stoichiometry of a series of K$_x$Fe$_{2-y}$Se$_2$ compounds. High quality single crystals were grown with the Bridgman technique, as described elsewhere in detail \cite{zhao1}. The transport properties are characterized by the in-plane resistivity which displays insulating, semiconducting, or superconducting behavior [Fig. \hyperref[fig1]{1(a)}]. These results are similar to those in previous reports \cite{donglai1}. The neutron scattering measurements were carried out on the HB-3A four-circle single-crystal diffractometer at the High-Flux Isotope Reactor at the Oak Ridge National Laboratory. The neutron wavelength employed was $1.000$ $\AA$ using a bent perfect Si-331 monochromator. No higher order neutron reflection is allowed with this experimental setup.  We define the wave vector $Q$ at ($q_x$, $q_y$, $q_z$) as ($h$, $k$, $l$)= ($q_xa/2\pi$, $q_ya/2\pi$, $q_zc/2\pi$) reciprocal lattice units in the orthorhombic unit cell to facilitate comparison with the iron pnictides. The neutron-diffraction data refinements are based on more than $200$ magnetic and nuclear reflections with the program FULLPROF \cite{fullprof}.

Fig. \hyperref[fig1]{1(b)} and \hyperref[fig1]{1(c)} display the magnetic and nuclear peaks associated with the $\sqrt{5} \times \sqrt{5}$ reconstruction at $5$ K in the insulating sample. These superstructure peaks can be fully described with the $\sqrt{5} \times \sqrt{5}$ block AFM order and iron vacancy order, consistent with the results reported previously \cite{weibao,ye,meng}. Our refinements suggest that this sample has a pure K$_2$Fe$_4$Se$_5$ (245) phase (or K$_{0.8}$Fe$_{1.6}$Se$_2$) with a magnetic moment $\sim 3.2(1)$ $\mu_B$ (Table \ref{table}).

\begin{figure*}[t]
\includegraphics[width=0.8\textwidth]{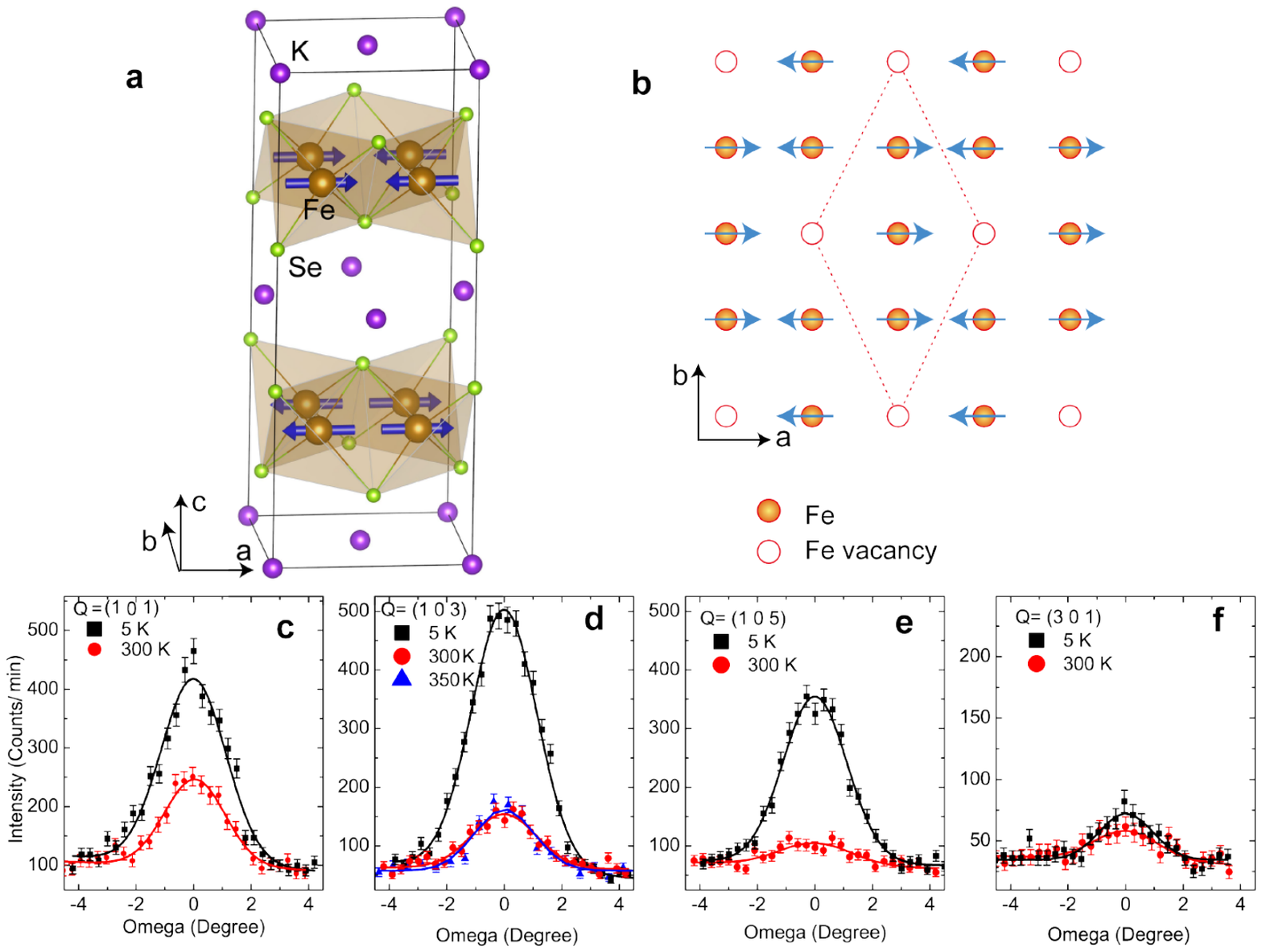}
\caption{\label{fig2}Stripe-type magnetic order and rhombus iron vacancy order of semiconducting K$_{0.85}$Fe$_{1.54}$Se$_2$. (a) The three-dimensional stripe-type magnetic structure of iron as determined from our neutron-diffraction data. (b) The in-plane magnetic structure and rhombus iron vacancy order. The red dashed line marks the 2 $\times$ 4 rhombus iron vacancy order. (c)-(f) A series of temperature-dependence magnetic and nuclear reflections associated with the stripe-type magnetic order and rhombus iron vacancy order. The extra intensities appearing at 5 K are associated with development of long range AFM order. The residual intensities at 300 K are due to rhombus iron vacancy order. (c) Omega scans near $Q$=(1 0 1); (d) $Q$=(1 0 3); (e) $Q$=(1 0 5); (f) $Q$=(3 0 1).
}
\end{figure*}

\begin{table*}
\caption{\label{table}Refined structural parameters of the 245 phase and the 122 phase in insulating, semiconducting, and superconducting K$_x$Fe$_{2-y}$Se$_2$}
\begin{ruledtabular}
\renewcommand\arraystretch{1.2}
\begin{tabular}[c]{lccccc}
& \multicolumn{2}{c}{Superconductor} & \multicolumn{2}{c}{Semiconductor} & Insulator \\
\hline
Average Composition & \multicolumn{2}{c}{K$_{0.85}$Fe$_{1.61}$Se$_2$} & \multicolumn{2}{l}{K$_{0.81}$Fe$_{1.58}$Se$_2$} & K$_{0.80}$Fe$_{1.60}$Se$_2$ \\
Phase & $122$ & $245$ & $122$ & $245$ & $245$ \\ \cline{2-3} \cline{4-5}
Phase Ratio & 16.7\% & 83.3\% &	25.2\% & 74.8\% & 1 \\
Moment $(\mu_B)$ & $0$ & $3.3(1)$ & $2.8(1)$ & $3.4(1)$	& $3.2(1)$ \\
K Occupancy & $0.88(6)$ & $0.84(7)$ & $0.85(7)$ & $0.80(6)$ & $0.80(7)$ \\
Fe Occupancy & $1.63(3)$ & $1.60(1)$ & $1.54(2)$ & $1.60(1)$ & $1.60(1)$ \\
Se Occupancy & $2$ & $2$ & $2$ & $2$ & $2$ \\
$R_1$ &	0.0334 & 0.0363 &0.0475 & 0.0555 & 0.0384 \\
$wRF^2$	& 0.0828 & 0.0763 &	0.0641 & 0.116 & 0.0719 \\
$\chi^2$ & 0.894 & 0.533 & 0.539 & 1.017 & 0.392 \\						
\end{tabular}
\end{ruledtabular}
\end{table*}

\begin{figure}[b]
\includegraphics[width=0.4\textwidth]{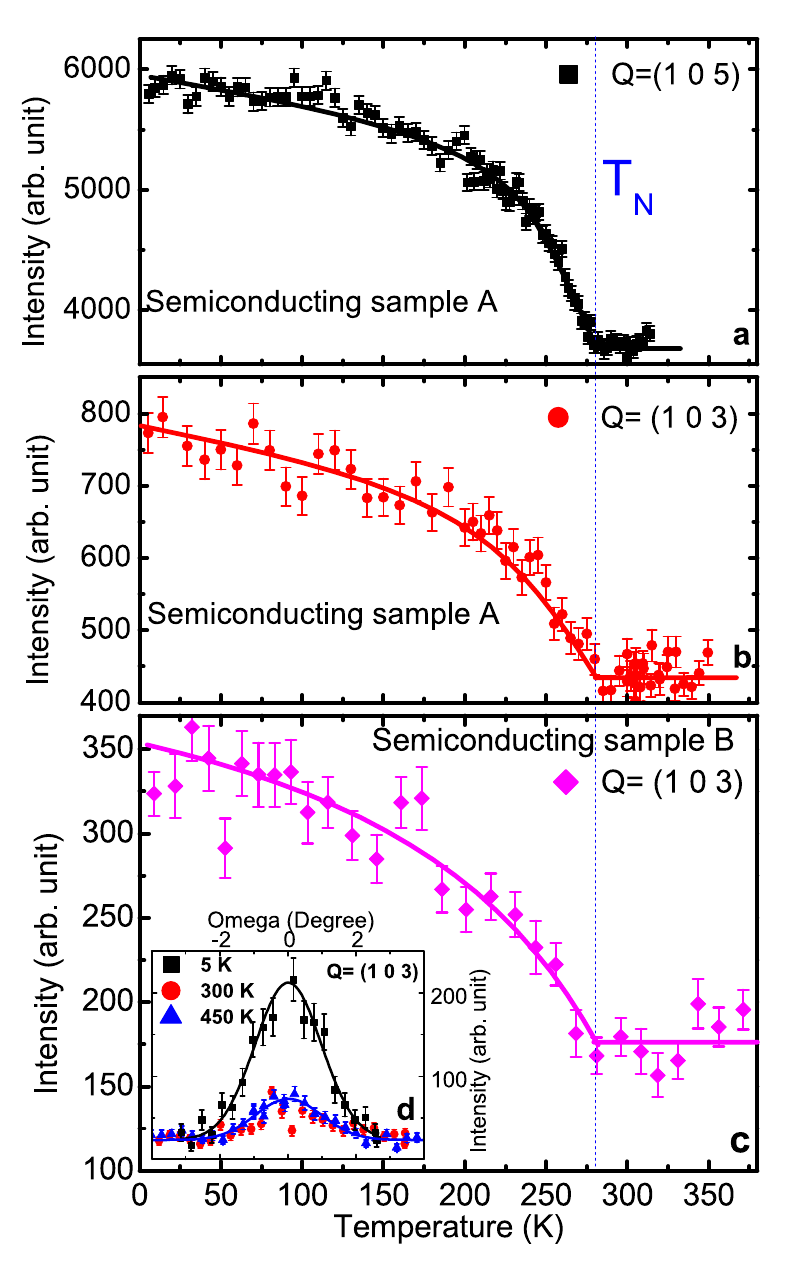}
\caption{\label{fig3}Magnetic order parameter in two different semiconducting K$_{0.85}$Fe$_{1.54}$Se$_2$ samples. The temperature dependence of the magnetic Bragg peak intensities displays a second order magnetic phase transition with the $\mathrm{N\acute{e}el}$ temperature $T_N$= $280$ K. (a) Temperature dependence of the $Q$=(1 0 5) magnetic peak of sample A. (b) Temperature dependence of the $Q$=(1 0 3) magnetic peak of sample A. (c) Temperature dependence of the $Q$=(1 0 3) magnetic peak of sample B. (d) Temperature dependence of the omega scans of $Q$=(1 0 3) peak of sample B. The peak exhibits little temperature dependence between 300 K and 450 K, indicating that the residual intensity above $280$ K is associated with the rhombus iron vacancy order.
}
\end{figure}

The most striking discovery is that in addition to the $\sqrt{5} \times \sqrt{5}$ superstructure peaks, we also observe in the semiconducting sample magnetic Bragg peaks at the $\sqrt{2} \times \sqrt{2}$ positions at 5K [Fig. \hyperref[fig2]{2(c)}-\hyperref[fig2]{2(f)}]. Here, $\sqrt{2} \times \sqrt{2}$ corresponds to (10$l$) in the orthorhombic unit cell shown in Fig. 2a. These peak intensities are significantly suppressed on warming to $300$ K, indicative of a magnetic phase transition [Fig. \hyperref[fig2]{2(c)}-\hyperref[fig2]{2(f)}].  The detailed temperature dependences of the $(1 0 3)$ and $(1 0 5)$ peaks show clear kinks at around $T_N$ = $280$ K, suggesting a second order phase transition [Fig. \hyperref[fig3]{3(a)} and \hyperref[fig3]{3(b)}]. Interestingly, some weak superstructure peaks are still present above $280$ K; these peaks exhibit negligible temperature dependence between $280$ K and $450$ K (Fig. \ref{fig2} and \ref{fig3}), demonstrating that the residual scattering is nuclear in origin. Our structual refinements suggest that the scattering above $280$ K could be associated with a rhombus iron vacancy order in the ThCr$_2$Si$_2$-type ($122$) phase as shown in Fig. \hyperref[fig2]{2(a)} and \hyperref[fig2]{2(b)}.

\begin{figure}[b]
\centering\includegraphics[width=0.46\textwidth]{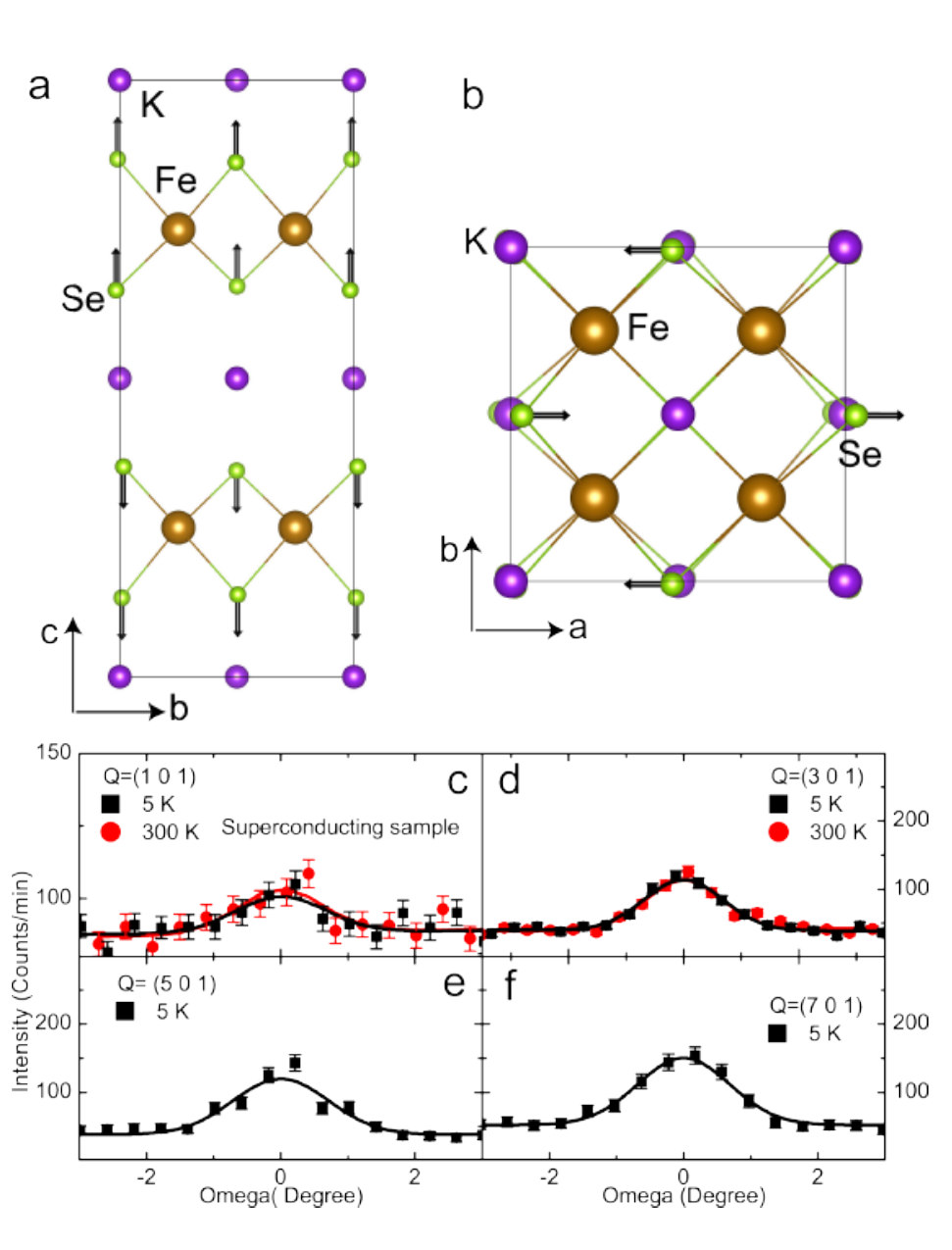}
\caption{\label{fig4} $\sqrt{2} \times \sqrt{2}$ Se distortion in the $122$ structure of the superconducting phase K$_{0.88}$Fe$_{1.63}$Se$_2$ ($T_c$= $30$ K). (a),(b) Schematic diagram of Se distortion; the arrows indicate the directions of the atom displacements with respect to perfect lattice. (a) Se distortion along the $c$ axis.  (b) Se distortion along the $a$ axis. (c)-(f) $\sqrt{2} \times \sqrt{2}$ reflections associated with Se distortion. The Bragg peak intensity increases as increasing $h$, suggesting it is nuclear in origin. (c) Omega scans of $Q$= (1 0 1) Bragg peak displays little temperature dependence between 5 K to 300 K, indicating that the stripe-type AFM order is completely suppressed. The weak scattering left is associated with Se distortion. (d) Temperature dependence of the $Q$= (3 0 1) Bragg peak. (e) The $Q$= (5 0 1) Bragg peak at 5 K. (f) The $Q$= (7 0 1) Bragg peak at 5 K.
}
\end{figure}

To determine the detailed magnetic structure in the 122 phase, we analyzed the integrated intensity of a series of magnetic Bragg reflections. The peak intensity increases from (1 0 1) to (1 0 3) and then decreases at (1 0 5), implying that the magnetic moments lie in the $ab$ plane, which is completely different from the $\sqrt{5} \times \sqrt{5}$ block AFM order with the moment along the $c$ axis. Refinements reveal that the magnetic peak intensities follow the Fe$^{2+}$ magnetic form factor and can be described with the same stripe-type magnetic structure as the iron pnictide parent compounds [Fig. \hyperref[fig2]{2(a)} and \hyperref[fig2]{2(b)}]. The Fe moment is antiferromagnetically ordered along the $a$ and $c$ axes and ferromagnetically ordered along the $b$ axis with the moment parallel to the $a$ axis. Here the $a$ axis is parallel to the short diagonal of the iron vacancy rhombus [Fig. \hyperref[fig2]{2(a)} and \hyperref[fig2]{2(b)}].

By taking into account the $\sqrt{5} \times \sqrt{5}$ iron vacancy ordered phase ($245$ phase) together with the $122$ phase with the stripe AFM order and the rhombus iron vacancy order discussed above, we can completely determine the magnetic and lattice structure parameters in the semiconducting sample (Table \ref{table}). Based on our refinements, the fractions of the $245$ phase and the $122$ phase in the semiconducting sample amount to $\sim$ $74.8\%$ and $\sim$ $25.2\%$, respectively. The refined stoichiometry of the 122 phase is K$_{0.85}$Fe$_{1.54}$Se$_2$. The iron moments of the stripe-type AFM order and block AFM order are  $2.8(1)$ $\mu_B$ and $3.4(1)$ $\mu_B$, respectively. We notice that the moment and stoichiometry of the block AFM order phase in the semiconducting sample are close to those of pure $245$ phase of the insulating sample, which is consistent with the fact that they have similar $\mathrm{N\acute{e}el}$ temperatures (Table \ref{table}) \cite{donglai1}. In a similar K$_x$Fe$_{2-y}$Se$_2$ sample, ARPES measurements also observed two different phases: an insulating phase with a $500$ meV gap and a semiconducting phase with a 40 meV gap \cite{donglai1}. Based on the above observations, we identify the stripe AFM ordered 122 phase as the semiconducting phase discussed earlier. Remarkably, a first principle calculation has predicted the essentially correct magnetic structure and iron vacancy order in a $122$-type structure in KFe$_{1.5}$Se$_2$ \cite{xiang}. The same calculation also suggests that this system is a band semiconductor with a $121$ meV gap \cite{xiang}. Therefore, we believe that the K$_{0.85}$Fe$_{1.54}$Se$_2$ phase with the $122$ structure discovered here is an antiferromagnetic band semiconductor, in contrast to the Mott insulating cuprates and metallic iron pnictides. We note that the refined potassium concentration deviates slightly from the commensurate KFe$_{1.5}$Se$_2$ phase, which could be due to the presence of disordered and localized vacancies or interstitials.

In addition to the results shown in Fig. \ref{fig2}, Fig. \hyperref[fig3]{3(a)} and \hyperref[fig3]{3(b)},  we also performed similar measurements in another semiconducting sample (semiconducting sample B). Very similar stripe-type AFM structure and magnetic order parameter temperature dependence are also observed [Fig. \hyperref[fig3]{3(c)}], demonstrating that the stripe-type AFM order is an intrinsic property of semiconducting samples.

We now discuss our results in the superconducting sample. Several STM measurements have suggested that superconducting samples contain both an insulating $\sqrt{5} \times \sqrt{5}$ iron vacancy order phase and a superconducting $122$-type phase with $\sqrt{2} \times \sqrt{2}$ superstructure which are spatially separated \cite{li2,wangyy,li}. Our refinements suggest that the fractions of $245$ phase and $122$ phase (refined stoichiometry K$_{0.88}$Fe$_{1.63}$Se$_2$) amount to $\sim$ $83.3\%$ and $\sim$ $16.7\%$, respectively (Table \ref{table}). Interestingly, while the block AFM order [refined moment $3.3(1)$  $\mu_B$] in this sample is still close to that of pure $245$ phase, the stripe-type AFM order and rhombus iron vacancy order are completely suppressed in the superconducting sample. Although some weak $\sqrt{2} \times \sqrt{2}$ reconstruction reflections are still present [Fig. \hyperref[fig4]{4(b)}-\hyperref[fig4]{4(f)}], their integrated intensities cannot be described with stripe AFM order or rhombus iron vacancy order. These weak reflections display negligible temperature dependences between $5$ K and $300$ K, indicating that they are nuclear in origin. Our refinement suggests that this $\sqrt{2} \times \sqrt{2}$ superstructure could be due to the Se distortion as illustrated in Fig. \hyperref[fig4]{4(a)} and \hyperref[fig4]{4(b)}. We note that the iron and potassium concentrations in the superconducting sample are higher than those of the 122 phase of the semiconducting sample, which naturally implies that the extra iron and potassium would introduce electron doping and thus suppress the stripe-type AFM order. The extra iron is also expected to suppress the rhombus iron vacancy order. Therefore, our findings suggest that this magnetic ordered semiconducting K$_{0.85}$Fe$_{1.54}$Se$_2$ is intimately related to the superconducting phase. Moreover, the magnetic moment of the $\sqrt{5} \times \sqrt{5}$ block AFM order in the $245$ phase is essentially unchanged in the insulating, semiconducting, and superconducting samples, implying that this phase is not directly relevant to the superconductivity (Table \ref{table}).


In summary, we have discovered a semiconducting antiferromagnetic phase (K$_{0.85}$Fe$_{1.54}$Se$_2$) with a rhombus iron vacancy order in the background of the $122$-type structure Fig. \hyperref[fig2]{2(a)} and \hyperref[fig2]{2(b)}]. The magnetic structure, iron vacancy structure, and magnetic moment ($\sim 2.8$ $\mu_B$) of K$_{0.85}$Fe$_{1.54}$Se$_2$ are consistent with first principle calculations that suggest a band semiconducting ground state in this system \cite{xiang}. Based on the fact that this new magnetic phase is completely suppressed in the superconducting sample as well as the explicit change in the composition in going from the semiconducting antiferromagnetic to superconducting sample, we conjecture that the superconducting state is derived from the newly discovered magnetic phase upon electron doping.
The discovery of a band semiconducting antiferromagnet bordering high $T_c$ superconductivity has profound implications. Clearly, a naive Fermi surface nesting picture  for the magnetism is not applicable for this system. Reference \cite{xiang} suggests that the semiconducting antiferromagnetism can be viewed at the atomic level due to the exchange interactions bridged by Se between Fe local moments. We also notice that the magnetic moment ($2.8$ $\mu_B$) and $\mathrm{N\acute{e}el}$ temperature ($280$ K) of semiconducting K$_{0.85}$Fe$_{1.54}$Se$_2$ are much higher than those of the 122 iron pnictides \cite{johnston}. This suggests that correlation physics plays a larger part in the magnetism for this system. The large magnetic moment in A$_x$Fe$_{2-y}$Se$_2$ may also explain why the magnetic resonance intensity in phase-separated Rb$_x$Fe$_{2-y}$Se$_2$ is comparable to that of the phase pure superconductor BaFe$_{2-x}$Ni$_x$As$_2$ \cite{keimer,chi}. The fact that this new phase exhibits the same magnetic structure as the parent compounds of other pnictides raises the possibility that the same magnetic fluctuations give rise to Cooper pairing, despite the mismatch of the magnetic wave vector and the fermiology in the normal state \cite{johnston,donglai,zhou,ding,donglai1}. In any case, the semiconducting antiferromagnet we discovered in this work interpolates between the antiferromagnetic Mott insulator of the cuprates and the antiferromagnetic semimetal of the iron pnictides. It might open a new window for finding new magnetic high $T_c$ superconductors.

We thank J.P. Hu for helpful discussions. This work is supported by the Director, Office of Science, Office of Basic Energy Sciences, U.S. Department of Energy, under Contract No. DE-AC02-05CH11231 and Office of Basic Energy Sciences US DOE DE-AC03-76SF008. J.Z. is supported by a fellowship from Miller Institute for Basic Research in Science. D.H.L is supported by DOE Grant No. DE-AC02-05CH11231. The research at Oak Ridge National Laboratory's High Flux Isotope Reactor is sponsored by the Scientific User Facilities Division, Office of Basic Energy Sciences, U. S. Department of Energy.


\end{document}